# Trigger-type metamaterials on the base of collective Jahn-Teller effect

*Eugene V. Rosenfeld*

Institute of Metal Physics, Yekaterinburg, Russia

Dr. E. V. Rosenfeld:  rosenfeld@imp.uran.ru

Generally, in case of the collective Jahn-Teller effect, a high-symmetry structure of a matrix in which quantum systems with degenerate ground state are inserted becomes distorted. This usually smooth transition can become abrupt only if the matrix by itself is a trigger and JTE merely activates its switching. It is shown in this paper that proper insertion into matrix of quantum systems with the singlet ground state and degenerate excited state leads to the formation of a new metastable state of the whole system and a stepwise appearance of JTE. A matrix of any nature can be transformed into trigger in this way if one manages to synthesize and insert into it proper quantum "active centers" with appropriate energy spectrum. Theoretically, this provides advanced possibilities for metamaterials development.

## 1. Introduction

In case of the classical Jahn-Teller effect (JTE),[1] a high-symmetry structure of a matrix in which quantum system (QS) with a degenerate ground state is inserted becomes distorted because the related growth of the matrix energy is compensated for by a decrease in the energy of the ground state of QS upon lifting the degeneracy. In this paper we will consider the matrices in which the number of implanted QSs is macroscopically large, so in what follows one should keep in view rather cooperative than classical JTE.[2] Moreover, it should be clearly understood that the matrix is not necessarily a crystal lattice and the distortion of the matrix structure does not necessarily mean the appearance of atomic shifts. In fact, it can be a classical system of any nature, for example, a magnetic system, see[3] and section 5 of this article. Of fundamental importance is only one: the distortion which leads to reduction of symmetry of the matrix must give rise to a field, which removes the degeneracy of the levels of the QSs.

In the absence of anharmonism the cooperative JTE results from a second-order phase transition (SOPT). However, if there is a metastable state with a distorted matrix structure, i.e., the matrix by itself is a trigger, a first-order transition (FOPT) can arise – JTE merely activates the switching of the already existing trigger.

In the work presented it is shown that proper insertion into any matrix of QSs with the singlet ground state and degenerate excited state can lead to the formation of a new metastable state of the whole system and, in such a case, JTE can result from FOPT. Practical application of this mechanism becomes possible in view of the fact that matrix of any nature can be transformed into trigger in this way if one manages to synthesize and insert into it appropriate quantum "active centers" with properly parameterized energy spectrum. Besides, since in this case upon emergence of JTE, a growth of the matrix energy is compensated for by a decrease in free energy of the ensemble of QSs, which is essentially related to the entropy growth, one can try to use this mechanism when designing refrigerating units.

Finally, the last remark necessary for a clear understanding of the results presented below. The order of the phase transition, first-FOPT when the matrix distortion proceeds stepwise or second- SOPT when it goes smoothly, is controlled by the dependence on the amplitude of distortion $d$ of two contributions to the free energy $F$ of the whole system – potential energy of matrix $E_m$ and free energy of QSs, which will be denoted as $F_{QS}$. However, taking into account that in the first approximation the splitting $2\delta$ of doublet in the spectrum of QS is always directly proportional to $d$, in what follows we assume that $F_{QS}$, $E_m$, and $F$ are mere functions of the splitting $\delta$.

## 2. QSs with degenerate ground state

In a general case, when anharmonic contributions to $E_m(\delta)$ are absent and the ground state of QS is doubly degenerate, the free energy of the system has the form:

$$F = F_{QS}(\delta) + E_m(\delta) = -k_B T \ln\left[2\cosh\left(\frac{\delta}{k_B T}\right)\right] + \frac{1}{2}r\delta^2. \tag{1}$$

Here, the parameter $r>0$ characterizes "generalized rigidity" of the matrix, *i.e.*, the increment of its energy (per one QS), upon distortion of the structure in the vicinity of equilibrium state. At weak distortions $\delta \ll k_B T$, free energy Equation (1) takes the form

$$F\big|_{\delta \ll k_B T} \approx -k_B T \ln 2 + \frac{1}{2}\left(r - \frac{1}{k_B T}\right)\delta^2 \tag{2}$$

where the contribution $-(2k_B T)^{-1}\delta^2$, square in $\delta$, from QSs to $F$ arises only because of a decrease in the ground-state energy upon splitting the doublet (expansion of the entropy contribution to $F_{QS}$ starts from the term $\sim \delta^4$). Thus, as a result of distortions, competitive contributions to the energy Equation (2) from the matrix and QSs appear and at low enough temperatures

$$T < T_C = (rk_B)^{-1} \tag{3}$$

the contribution of QS prevails, the distortion becomes energetically favorable, and JTE arises.

At strong distortions $\delta \gg k_B T$, the contribution of QSs to $F$ becomes nearly linear, so that the sum of energies of the matrix and QSs

$$F \approx -\delta + \frac{1}{2}r\delta^2 \tag{4}$$

necessarily has a minimum at $\delta_{min} = r^{-1}$. Combining conditions $\delta \gg k_B T$ and $\delta_{min} = r^{-1}$ one obtains $r^{-1} \gg k_B T$ *i.e.* $T \ll T_C$ and so this minimum actually corresponds to the ground state only at low temperatures.

The exact position of extremes of $F$ Equation (1) for any $T$ can be obtained from the equation $\partial F/\partial \delta = 0$ which has two roots. The first of them is trivial, $\delta$=0, and always exists, whereas the second is defined by the root of well-known equation

$$\frac{T}{T_C} = \frac{\tanh(x)}{x}, \quad x = \frac{\delta}{k_B T}. \tag{5}$$

Since the right hand side of this equation monotonously decreases from 1 to 0 with growing *x*, its root exists only at $T < T_C$. With decreasing temperature, this root corresponding to the minimum of *F* arises at a point $\delta$=0 at the moment when $T = T_C$, and the already existing minimum transforms into maximum at this point, see Equation (2). Upon the further lowering the temperature, this root gradually shifts to the region $\delta \neq 0$, so that the phase transition in the system with free energy Equation (1) must necessarily be SOPT. Consequently, to transfer it into FOPT, one has to either retard the disappearance of the minimum at $\delta=0$ or accelerate the appearance of the minimum at $\delta \neq 0$.

It is clear that since the contribution of QS at $\delta \gg k_B T$ is virtually linear $F_{QS} \approx -\delta$, see Equation (4), an «untimely» appearance of the second minimum of $F$ in the region of large $\delta$ can be related to solely anharmonic contributions to the matrix energy:

$$E_m(\delta) = \frac{1}{2} r\delta^2 - \frac{1}{4} p\delta^4 + \frac{1}{6} q\delta^6 + ...; \quad p, q > 0. \tag{6}$$

If $p^2 > 4rq$, this function has, along with the minimum at $\delta=0$, a minimum at $|\delta| \neq 0$, and owing to linear contribution from QS, this local minimum of $E_m(\delta)$ can transform into a global minimum of $F(\delta)$ when the minimum at $\delta = 0$ is still present, see **Figure 1a**. Thus, the conventional JTE can arise as a result of FOPT only if the matrix has two different equilibrium states, i.e., is by itself a trigger and JTE merely launches its switching. It should be noted, however, that a softer variant is possible as well, when a metastable state of the matrix is only in the process of origination (emerging kink of $E_m(\delta)$, see Fig.ure1a).

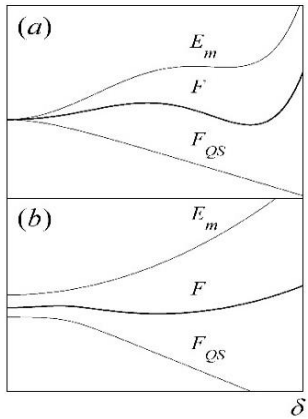

**Figure 1**. Two minima in the dependence of free energy of a system $F = F_{QS} + E_m$ on splitting of the doublet $\delta$ can arise either (*a*): because of anharmonic contributions Equation (6) to the matrix energy that condition the existence of a local minimum or inflection in the curve $E_m(\delta)$ or (*b*): because at low $\delta$, the contributions of QS to the «rigidity» of the system (and related curvature of the line $F_{QS}$) falls as a consequence of low occupancy of the doublet in the exited state, see Equation (8).

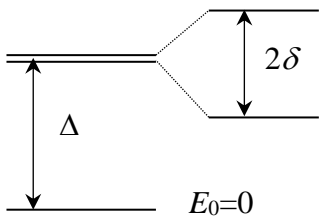

**Figure 2**. Splitting of the upper doublet level upon the matrix deformation.

### 3. QSs with a singlet ground and degenerate excited state

Another mechanism of the emergence of FOPT can realize itself even in the absence of anharmonism in the matrix when its energy has the only minimum. This takes place if the force that QS applies to the matrix, "pushing" it still farther from the equilibrium, falls dramatically in the region of weak distortions, i.e., the slope of $F_{QS}$ curve will be far less at low $\delta \ll k_B T$, see. Figure 1b. It is just the way how the three-level QS with a singlet ground state and doublet excited one (see **Figure 2**) behaves. Now instead of Equation (1) we obtain

$$F = \Delta - k_B T \ln\left[\exp\left(\frac{\Delta}{k_B T}\right) + 2\cosh\left(\frac{\delta}{k_B T}\right)\right] + \frac{1}{2}r\delta^2, \qquad (7)$$

and the roles that the energy and the entropy play in the course of JTE emergence change.

### 3.1. The role of entropy

When distortions are small, $\delta < \Delta$, the average energy of QS grows rather than lowers with $\delta$, since the occupancy of the excited level with the energy $\Delta - \delta$ increases. However, it is the reason for which the entropy grows simultaneously, and so $F_{QS}$ lowers:

$$\langle E_{QS} \rangle \approx (\Delta - \delta)\exp\left(\frac{\delta - \Delta}{k_B T}\right), \quad S \approx \frac{\Delta - \delta + k_B T}{T}\exp\left(\frac{\delta - \Delta}{k_B T}\right), \quad F_{QS} \approx -k_B T \exp\left(\frac{\delta - \Delta}{k_B T}\right) \qquad (8)$$

At low temperatures $k_B T \ll \Delta$ and small deformations of the matrix $\delta \ll \Delta$, this contribution to $F$ turns out exponentially small, so that the maximum of $F_{QS}$ in Figire 1b transforms into a plateau. Therefore, the respond of the system to small deformation is controlled only by the matrix rigidity – the $F$ minimum at $\delta=0$ becomes stabilized. At the same time, with growing $\delta \to \Delta$, the value of the negative contribution $F_{QS}$ grows first exponentially and then linearly, which results in the appearance of the $F$ minimum even at smooth dependence $E_m(\delta)$, see Equation (4) and Figure1b. It is important to note that this decrease of $F_{QS}$ is caused by the growth of the entropy rather than lowering of the internal energy of QSs, at least as long as $\delta < \Delta$. Thus, FOPT arises as the distorted state becomes thermodynamically more probable rather than because its lower energy.

Note also that since upon growing the splitting in the range $\delta < \Delta$, an increasing number of QSs turn out to be thrown from the ground to the lowest of the excited state at the expense of the thermal energy of the matrix, the appearance of JTE should result in a decrease the temperature of the heat-isolated system (QSs+matrix), see[4] in this context.

## 4. Phase diagram

It is more pertinent to conduct an analysis of the phase diagram of the system via studying extremes of the more general than Equation (1) and Equation (7) function

$$\Phi(z) = -\ln[a + 2\cosh(z)] + \frac{1}{2}bz^2, \quad z = \frac{\delta}{k_B T}, \quad a = \exp\left(\frac{\Delta}{k_B T}\right), \quad b = rk_B T, \quad (9)$$

which, depending on the form of $a$, describes systems with two, three, and four levels. The exact position of nontrivial extremes of $\Phi$ is controlled by the roots of the equation which we write, to underline its similarity to Equation (5), in the following form

$$\frac{a+2}{2}b \equiv \frac{T}{T_{eff}} = f(z) \equiv \frac{\tanh(z)}{z} \cdot \frac{(a+2)\cosh(z)}{a+2\cosh(z)}, \quad T_{eff} = \frac{2}{(a+2)rk_B}. \quad (10)$$

It is easily seen that at any $a < 4$ (and, in particular, at $a=0$, when Equation (10) transforms into Equation (5)), function $f(z)$ in Equation (10), similarly to the right hand side of equation (5), monotonously decreases from 1 to 0 with growing $z$, so that in the temperature range $T < T_{eff}$, this equation has the only real root, corresponding to the minimum of $\Phi$. However, in the range of $a > 4$,[5] in the plot $f(z)$ there appears a maximum (see **Figure 3**) at the point $z_{max}$ which is the root of equation

$$\frac{2 + a\cosh(z)}{a + 2\cosh(z)} = \frac{\sinh(z)}{z}. \quad (11)$$

Now, in the temperature range $T_{eff} < T < T_{eff} f(z_{max})$ Equation (10) has two roots, the lower one corresponding to the maximum and the higher, to the minimum of $\Phi(z)$. Consequently, in this range of values of $a$ and $b$, function $\Phi(z)$ has two minima and, when changing the parameters, the matrix distortion should arise and vanish as a result of FOPT.

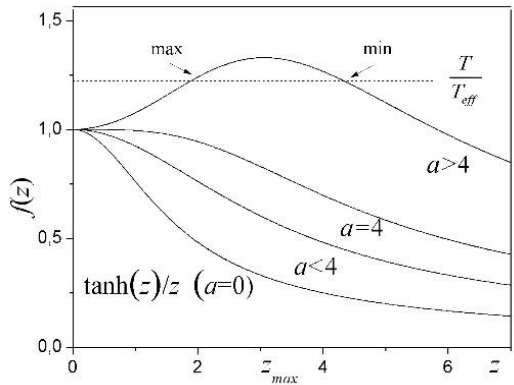

**Figure 3**. Form of function $f(z)$ Equation (10) at different $a$ values. At $T > T_{eff} \cdot f(z_{max})$, the deformed state of the matrix is unstable. At $T < T_{eff}$, on the contrary, only the deformed state is stable. At $T_{eff} < T < T_{eff} \cdot f(z_{max})$, one of the states is stable, whereas the other, metastable – this is the region of FOPT.

A simple phase diagram of the system with two lines sectioning the plane of $a, b$ parameters into three regions is shown in **Figure 4**. In region 1 below the line $b = 2(a+2)^{-1}$ only the deformed state of the matrix is stable. At $a=4$, this line intersects with the line $b = 2(a+2)^{-1} f(z_{max})$ and in region 2 between these two lines two minima of $\Phi$ coexist and the FOPT is possible. At last, in region 0 only the undistorted state of the matrix is stable.

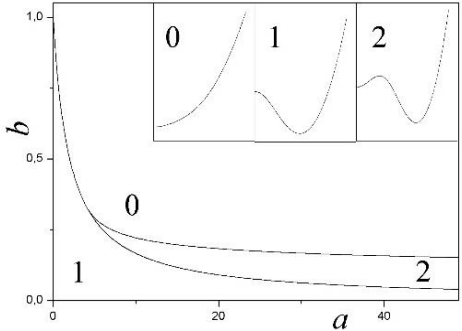

**Figure** 4. Phase diagram of a system with free energy $k_B T \Phi$ Eqiation (9). Function $\Phi$ has the only minimum in region 0 - at $z=0$ and in region 1 - at $z \neq 0$, whereas in region 2, two minima (form of $\Phi$ in each region is shown in the corresponding insets).

## 5. Example of real physical system

Thus, QSs with the singlet ground state and doublet excited state are active centers, which, being inserted into any matrix in a sufficient amount, can turn it into a trigger. It is natural to expect that this effect should be given rise by active centers with a more complicated spectrum of excited states as well. As an example of a four-level QS of such type, let us consider a pair of antiferromagnetically coupled spins ½ with Hamiltonian $J(\hat{\mathbf{s}}_1 \cdot \hat{\mathbf{s}}_2)$, $J > 0$. In this system, the state with the total spin of the pair $S = 1$ will be separated from the state with $S = 0$ by the gap $\Delta = J$, and under action of external or exchange field, only states with $S^z = \pm 1$ will be split.

Let us take as a matrix a couple of classical moments $\mathbf{M}_1$ and $\mathbf{M}_2$ that are antiferoomagnetically coupled and create a field acting upon QS. Then, Hamiltonian of the whole system takes the form

$$\hat{H} = +J(\hat{\mathbf{s}}_1 \cdot \hat{\mathbf{s}}_2) - h(\mathbf{M}_1 + \mathbf{M}_2)(\hat{\mathbf{s}}_1 + \hat{\mathbf{s}}_2) + A(\mathbf{M}_1 \cdot \mathbf{M}_2), \quad A > 0. \quad (12)$$

As an appropriate object for application of this model one can choose compounds of the type RFe$_x$Al$_{12-x}$ [6], where spins of Fe atoms occupying sites inside each 8f-plane of the ThMn$_{12}$ –type lattice order ferromagnetically forming large classical moments $\mathbf{M}$. Sites in the 8j-planes that lie between the 8f-planes are randomly occupied by atoms of Fe and Al, and Fe atoms with spins $\hat{\mathbf{s}}$ can be grouped there into anferromagnetic couples.

If $\mathbf{s}_1$ and $\mathbf{s}_2$ are taken as classical vectors, then energy Equatipn (12) turns out the squared form of cosines of angles $\theta$ and $\Theta$, where $2\theta$ is the angle between $\mathbf{s}_1$ and $\mathbf{s}_2$, and $2\Theta$ is the angle between $\mathbf{M}_1$ and $\mathbf{M}_2$:

$$E = const + 2Js^2 \cos^2\theta - 4hsM\cos\theta\cos\Theta + 2AM^2\cos^2\Theta. \tag{13}$$

This energy has the only minimum, the position of which (in case of a canted structure) will change smoothly with changing the parameters or external magnetic field. If to take into account that $\hat{\mathbf{s}}_1$ and $\hat{\mathbf{s}}_2$ are quantum operators, then the free energy corresponding to Equation (12) is equal to

$$F = \Delta - k_B T \ln\left[\exp\left(\frac{\Delta}{k_B T}\right) + 1 + 2\cosh\left(\frac{2hM\cos\Theta}{k_B T}\right)\right] + 2AM^2\cos^2\Theta. \tag{14}$$

At $\delta = 2hM\cos\Theta$, this expression differs from Equation (7) by substitution $\exp(\Delta/k_B T) \to 1 + \exp(\Delta/k_B T)$, so that at the corresponding parameter values, canting of moments $\mathbf{M}_1$ and $\mathbf{M}_2$ (i.e., metamagnetic transition) should take place stepwise. it is not improbable that similar mechanism can give rise to metamagnetic transitions in magnetics of another type as well.

## 6. Conclusion

We see that a possibility of excitation of jumplike Jahn-Teller transition in a matrix of any nature exists conceptually. Nevertheless, in general to implement this idea in practice one should solve the problem of immense complexity related to creation of the active center.[7] The energy spectrum of this built in the matrix QS should meet the following requirements:

- The ground state is singlet.
- The first excited state is degenerate when the matrix is in the highly symmetrical configuration.
- The degeneracy must be lifted when the distortion arises (it is highly plausible that some kind of amplifier or field converter should be embedded in the system at this stage).
- The width of the gap above the ground state must have appropriate value.

If all these problems are solved, it will only be necessary to integrate a sufficient number of QSs into the matrix and launch the transition, varying external field, temperature etc.